\def \yskip{\penalty-50\vskip3pt plus 3pt minus 2pt}
\def \reference{\par \yskip \noindent \hangindent .4in \hangafter 1}
\def \abc#1#2#3#4 {\reference#1, {\sl#2}, {\bf#3}, #4}
\def \blank {\lower 5pt\hbox to 0.75in{\hrulefill}}

\def \cm{~\rm{cm}}
\def \s{~\rm{s}}
\def \km{~\rm{km}}

\def \AU{~\rm{AU}}
\def \yrs{~\rm{yrs}}
\def \yr{~\rm{yr}}
\def \K{~\rm{K}}
\def \G{~\rm{G}}
\def \erg{~\rm{erg}}
\def \kpc{~\rm{kpc}}
\def \lae{\mathrel{<\kern-1.0em\lower0.9ex\hbox{$\sim$}}}
\def \gae{\mathrel{>\kern-1.0em\lower0.9ex\hbox{$\sim$}}}

\documentstyle[11pt,paspconf]{article}

\begin{document}
\noindent To appear in the Proceedings of the {\it 10th Cambridge Workshop
on Cool Stars, Stellar Systems and the Sun}, July 15-19, 1997. 

\title{THE NUMBER OF PLANETS AROUND STARS}
\author{Noam Soker}
\affil{Department of Physics, University of Haifa at Oranim, 
Tivon 36006, Israel; soker@phys1.technion.ac.il}

\begin{abstract}
 Based on the large number of elliptical planetary nebulae I argue that 
$\sim 55 \%$ of all progenitors of planetary nebulae have planets around 
them. 
 The planets spin up the stars when the later evolve along the red giant 
branch or along the asymptotic giant branch. 
 The arguments, which were presented in several of my earlier works, 
and are summarized in the paper, suggest that the presence of four 
gas-giant planets in the solar system is the generality rather than 
the exception.   
 I continue along this line and study two aspects of the star-planet 
interaction paradigm:
 (1) I examine the possibility of detecting signatures of surviving 
Saturn-like planets inside planetary nebulae.  
 (2) I propose a model by which the second parameter of the 
horizontal branch, which determines the distribution of horizontal 
branch stars in the HR diagram, is the presence of planets. 
 A red giant branch star that interacts with a planet will lose a large 
fraction of its envelope and will become a blue horizontal branch star.

\end{abstract}
\bigskip

\noindent {\bf preprints cited in the text can be obtain by sending 
a request to soker@phys1.technion.ac.il}


\section{Introduction}
\subsection{Axisymmetrical Planetary Nebulae: General} 
Almost all well resolved planetary nebulae (PNs) images deviate substantially 
from sphericity, having rather large-scale elliptical or bipolar shapes. 
  I follow Schwarz, Corradi \& Stanghellini (1993; see also 
Corradi \& Schwarz 1995), in referring to more or less  
axisymmetric PNs which have two lobes and a morphological
``waist'' between them as ``bipolar PNs'' (these PNs are also called 
``butterfly'' or ``bilobal'').  
PNs which have a more elliptical than bilobal structure, i.e., have
no morphological waist, are termed elliptical PNs.
 This well-known typical structure of most PNs led to a two-decade-old 
debate on whether elliptical PNs can be formed through 
single-stellar evolution, or whether a binary companion is necessary 
(Soker 1997 and references therein). 
 
 In my opinion the binary system paradigm is doing overwhelmingly 
better than the single stellar evolution paradigm.  
 First, models which try to explain the axisymmtrical structures of
PNs encounter {\it severe} problems, since they need to assume angular 
velocity which are impossible for singly evolved stars to acquire.  
 This was demonstrated by Soker (1996b) for one model, and it is 
shown in the appendix below for two other models. 
 As it stands now, all models for axisymmtrical mass loss on the AGB
or beyond require the envelopes of the stars to be spun-up by  
binary companions, being stellar or substellar. 
 
 Second, in two recent papers I compared the basic properties of 
singly-evolved stellar models with those of binary models. 
 In the first paper of the two (Soker 1997; $\S 1.2$ below summarizes
the second paper) I listed 9 key observations 
that any model should comply with and explain.  
 I found that binary based models are doing much better than singly-evolved
stellar models do.  
 In most cases the role of the binary companion is the 
{\it deposition of angular momentum}, through tidal force or common 
envelope evolution. 
 This can be done by stellar or substellar (brown dwarfs or planets) 
companions.
 I proposed four evolutionary classes, and examined the morphologies 
of 458 PNs to find the prevalence of each class: 
\newline
(a) Progenitors of PNs which did not
interact with any companion, and therefore form spherical PNs 
(not considering interaction with the ISM and small scale structures). 
These amount to $\sim 10 \%$ of all PNs.
\newline
(b) Progenitors which interact with stellar companions which 
avoided the common envelope phase for a large portion of the
interaction time. 
These form bipolar PNs (Corradi \& Schwarz 1995)  
and amount to  $11 ^{+ 2}_{-3} \%$ of all PNs.
\newline
(c) Progenitors which interact with stellar companions via common envelope
phase, $23^{+11}_{-5} \%$ of all nebulae.  
These are mainly elliptical with high concentration of mass in the
equatorial plane (Bond \& Livio 1990; Pollacco \& Bell 1997). 
In some cases they form bipolar PNs (NGC 2346) 
\newline
(d) Progenitors which interact with {\it substellar} (i.e. planets and brown 
dwarfs) companions via common envelope phase, $56^{+5}_{-8} \%$ 
of all nebulae.  These form elliptical PNs with moderate density
contrast between the polar directions and equatorial plane.

\subsection{Bipolar Planetary Nebulae} 
 In a recent paper (Soker 1998) I concentrated on bipolar PNs, 
which amount to $\sim 11\%$ of all PNs, and which are formed from
massive, $M \gae 1.5 M_\odot$ progenitors (Schwarz \& Corradi 1995; 
Stanghellini 1995).  
 That bipolar PNs are formed from massive progenitors was the main,
and sometimes the only, reason for many researchers to argue that 
bipolar PNs are formed from singly-evolved stars. 
 In that paper I listed ten critical observations, and argued that 
single star models for the formation of bipolar PNs have difficulties 
in complying with several of these observations.
 On the other hand, binary system progenitors can naturally explain
these key observations, and in addition explain the rich varieties  
of structures possessed by bipolar PNs
(e.g., Harpaz, Rappaport \& Soker 1997; Soker Harpaz \& Rappaport 1998).
 Based on three of the critical observations, on several works 
by Corradi \& Schwarz (1995 and references therein) and on the scenario  
proposed by Morris (1987), I postulated that the progenitors of bipolar 
PNs are binary stellar systems in which the secondary diverts a substantial 
fraction of the mass lost by the asymptotic giant branch primary, 
but the systems avoid the common envelope phase for a large fraction of 
the interaction time. 
 This scenario predicts that the central stars of most bipolar PNs 
are in binary systems having orbital periods in the range of 
a few days to few$\times 10 \yrs$. 
 
 I proposed also (Soker 1998) an explanation for the positive correlation 
of bipolar PNs with massive progenitors in the paradigm of 
binary system progenitors. 
 I suggested that the main difference between massive ($M \gae 2 M_\odot$)
and low mass progenitors is the larger radii which low mass stars attain 
on the red giant branch (RGB). 
 These larger radii on the RGB cause most stellar binary companions, 
which potentially could have formed bipolar PNs if the primary had been
on the AGB, to interact with low mass primaries already on the RGB. 
These systems may enter a common envelope phase on the RGB.
   In some cases the strong interaction will cause the primary to lose most,
or even all, its envelope on the RGB; such systems will form only faint PNs, 
or no PN at all. 
 In other cases of common envelope evolution the secondary will spiral-in,
but the primary will retain its envelope, and eventually evolve to
the AGB phase. 
 Such systems will enter a common envelope phase early on the AGB, and
form elliptical, rather than bipolar, PNs. 
  Massive primaries on the other hand, reach large radii only on the AGB,
and therefore their companions interact mainly on the AGB,
 the stage prior to the PN phase. 
 In addition, more massive primaries retain much more massive envelopes,
which result in high density concentration in the equatorial plane of the
descendant PNs. 
 
\subsection{The Number of Planets Around Stars} 

 The postulate that most elliptical PNs result from the influence
of substellar objects, mainly gas giant planets, lead to the conclusion 
that planets are commonly present around main sequence stars. 
 To get a more quantitative estimate, I derived the maximal orbital 
separations allowed for brown dwarfs and massive planets in order 
to tidally spin-up progenitors of PNs (Soker 1996a). 
 I found the maximal orbital separation to be $\sim 5~$AU.  
For a substellar object to have a high probability of being present 
within this orbital radius, on average several substellar objects must 
be present around most main sequence stars of masses $\lae 5 M_\odot$. 
 As stated in the first subsection, according to the star-planet
interaction paradigm $\sim 55 \%$ of all main sequence stars
which are progenitors of PNs should have such planetary systems. 
 My arguments suggest that the presence of four gas-giant planets
in the solar system is the generality rather than the exception.   

 In the next two sections I will examine other aspects of the presence of 
planetary systems: The possibility to detects the surviving outer planets 
in elliptical PNs, and the possibility that the presence of planets
is the ``second parameter'' which determines the morphology of the
horizontal branch on the HR-diagram. 

\section{Detecting Planetary Systems inside Planetary Nebulae} 

 According to the binary model paradigm which was presented in the 
previous section, up to several Saturn-like planets are being present 
around the central stars of many elliptical and spherical PNs. 
 The orbital separation of these surviving planets will be 
$a \gae 5 AU$.  
 As I now suggest, these planets can be detected during the PN phase. 
 Two factors make the planets more likely to be detected during the PN 
phase: 
the high luminosity of the central star and its energetic wind. 
 That planets can reveal themselves around evolved stars was suggested before.
 Struck-Marcell (1988) proposed that SiO masers in Mira stars may originate 
in the magnetospheres of gas-giant planets. 
 This would require, as noted already by Struck-Marcell (1988),
that several planets are present around many sun-like stars. 

 Dopita \& Liebert (1989; hereafter DL89) proposed that the unresolved 
compact nebula around the central star of the PN EGB 6 results from the 
ablation of a Jovian planet. 
 I will follow the general idea of DL89, but will detour as necessary.
DL89 assume that the planet's distance from the central star is $2-4 \AU$. 
 Because of tidal effects (Soker 1996a) 
I do not expect planets at such close orbital separations to survive the 
primary AGB phase, and therefore I take $a \gae 5 \AU$. 
 DL89 conclude that the effect of the wind is small compared with that 
of the ionization radiation.  
 I will try to show that the winds, both during the AGB and the PN phase, 
have very interesting effects.  
 I start by estimating the accretion rate during the AGB phase.
 The neutral material blown by the AGB phase will not be influenced much
by the planet's magnetic field, whereas the ionized gas may be captured by
the field. 
 For the neutral material, the accretion radius is 
$R_A = R_p (v_{\rm {esc}}/v_r)^2$, where $R_p$ is the planet's radius,
$v_{\rm {esc}}$ is the escape velocity from the planet's surface, and
$v_r=(v_w^2 + v_O^2)^{1/2}$ is the planet relative velocity to the wind. 
 I have neglected the sound speed in the cool wind, since it is 
smaller than both the wind velocity $v_w$ and the orbital velocity $v_O$. 
 Taking a Jupiter like planet and a typical AGB star, we find,  
$v_r \simeq 15 \km \s^{-1}$, $v_{\rm {esc}} \simeq 50 \km \s^{-1}$, 
and therefore $R_A \simeq 1 R\odot$. 
 The mass fraction of the AGB wind which is accreted by the planet is
$f_a = (R_A/2 a)^2 \simeq 10^{-7} (a/10 \AU)^{-2}$. 
 For a central star envelope's mass of $1 M_\odot$, the total accreted mass
onto the planet is $M_{\rm {acc}} \simeq 10^{-7} M_\odot$.   
 Because of the high luminosity of the central star, and the heat
released by the accretion, this accreted material will stay at a 
temperature of $\sim 10^3 \K$. 
 The importance of this accreted layer of relatively hot gas is that
it will be the first to be lost during the PN phase. 

 During the PN phase the wind is hot and ionized, and it deposits energy
into the planet's magnetosphere.  
 This will heat the atmosphere and lead to the emission of auroral lines. 
  A typical fast wind of PNs' central stars  
($\dot M_{fw} \sim 10^{-7} M_\odot \yr ^{-1}$ and 
$v_{fw} \sim 1,000 \km \s^{-1}$)
is $\sim 10^7$ more energetic than the solar wind. 
 To avoid the complicated calculations of the physics of the auroral lines 
and the size of the magnetosphere, I scale the intensity with the
wind energy. 
 Taking typical auroral line intensities of Jupiter 
(e.g., Kim, Caldwell, \& Fox 1995) I find that the flux from a planet 
inside a PN at a distance of $1 \kpc$ is 
$<10^{-19} \erg \cm^{-2} \s^{-1} \AA^{-1}$, which is bellow current 
detection limits. 
  
 For a central star luminosity of $L = 1,000 L_\odot$, the ablation rate
estimated by DL89, to an order of magnitude, is 
$\dot M_p \sim 10^{-12} M_\odot \yr^{-1}$.  
  In estimating the stripping rate caused by the fast wind, DL89 did not
consider the magnetosphere, which increases the cross section for the
interaction of the fast wind with the planet. 
 The planet's magnetic field pressure is given by 
$P_B = B_s ^2 (r/R_p)^{-6}/8 \pi$, where $B_s$ is the magnetic surface field, 
and I am assuming a dipole magnetic field. 
 Equating this pressure to the ram pressure of the fast wind yield the 
magnetosphere radius $R_m$
\begin{equation}
R_m =  5 R_p
\left( {{B_s}\over{10 \G}} \right) ^{1/3}
\left( {{a}\over{10 \AU}} \right) ^{1/3}
\left( {{\dot M_{fw}}\over{10^{-8}}} \right) ^{-1/6}
\left( {{v_{fw}}\over{1,000}} \right) ^{-1/6},
\end{equation}
where mass loss is given in $M_\odot \yr^{-1}$ and velocity in $\km \s^{-1}$.
 I follow DL89, and equate the momentum of the fast wind, but now the
momentum impinging on the magnetosphere, to that of the stripped material.
 DL89 determine the stripping rate, whereas I am interested in the velocity
which the fast wind can impart to the tail of the ablated material: 
\begin{equation}
v_{tail} \simeq 0.5  
\left( {{\dot M_p}\over{10^{-12}}} \right)^{-1}
\left( {{R_m}\over{1 R_\odot}} \right)^{2}
\left( {{a}\over{10 \AU}} \right)^{-2} 
\left( {{\dot M_{fw}}\over{10^{-8} }} \right) 
\left( {{v_{fw}}\over{1,000}} \right)  \km \s^{-1},
\end{equation}
where units are as in the previous equation. 
 To be accelerated to a velocity of $>10 \km \s^{-1}$ by the fast wind 
the tail should be extended to a distance of $\sim 10 R_m$.
The crossing time of this distance is $\sim 0.1 \yr$. 
 To reach a velocity of $\sim 100 \km \s^{-1}$ the tail will 
probably extend to a distance of $\gae 10^{13} \cm$, with a crossing
time of $\gae 1 \yr$. 
  The planets will move only a few percents of its orbit during that time. 
To estimate the total number density in the tail I equate its thermal 
pressure to that of the fast wind's ram pressure, assuming the tail is mostly 
ionized and at a temperature of $10^4 \K$.
 This gives
\begin{equation}
n \simeq 10^8 
\left( {{a}\over{10 \AU}} \right) ^{-2}
\left( {{\dot M_{fw}}\over{10^{-8}}} \right) 
\left( {{\dot v_{fw}}\over{1,000}} \right) \cm ^{-3},
\end{equation}
where units are as in the previous 2 equations. 
 The total mass accumulated during a year will be $\sim 10^{-12} M_\odot$,  
and therefore the volume this material occupies is $\sim 10^{37} \cm^3$. 
 This nebula is $\sim 30$ times denser than that of EGB 6 (DL89), 
and $\sim 10^4$ smaller, therefore, it is $\sim 10-100$ fainter. 
 As the nebulae evolves, it will mix with the hot fast wind 
material and dispersed, it may become similar to the nebula of EGB 6. 
 A plausible conclusion of the discussion above is that planets around the
central stars of PNs {\it may} reveal themselves as compact (and unresolved)
nebulae around the central stars. 
 The velocity of the nebula will change on a period of several tens years
with an amplitude of $\sim 10 \km \s^{-1}$.  
 I should stress that the discussion above is based on estimates, and not 
on detailed calculations. 
 Therefore, it might be that the compact nebula will be fainter than what 
I found above, and below detection limits. 
 In any case, I strongly encourage careful observations of unresolved 
emission nebula around the central stars of PNs. 

\section{The Second Parameter of the Horizontal Branch: Planets?}
 
 The second parameter problem, which is more than thirty years old 
(for recent reviews see Rood 1997; Rood Whitney \& D'Cruz 1997),
is the question of the physical process that determines the 
color-magnitude distribution of stars on the horizontal branch (HB). 
 In a number of globular clusters the HB extends
toward the blue side, i.e., high effective temperatures. 
 This region is termed the blue HB. 
 In a few clusters there is a bimodal distribution of red and blue
HB stars. 
 The distribution in the HR diagram requires that the blue HB stars lose
up to almost all their envelope while on the RGB (Dorman, Rood \& O'Connell 
1993; D'Cruz {\it et al.} 1996 and references therein). 
D'Cruz {\it et al.} (1996) show that they can reproduce the basic 
morphology of the HB in different globular clusters by assuming a simple 
mass loss  behavior on the RGB.
 They could even produce the bimodal distribution for solar metalicity
clusters.    
 However, I think that there are still open questions. 
\newline
{\bf (1)}  The bimodal distribution is found 
in low metalicity globular cluster as well. 
\newline 
{\bf (2)} What determines the distribution of the mass 
loss rates on the RGB?
\newline 
{\bf (3)} How come HB stars have rotation velocities of up to 
$\sim 40 \km \s^{-1}$ (Peterson, Rood \& Crocker 1995)?
Harpaz \& Soker (1994) show that the envelope's angular
momentum of evolved stars decreases with mass loss as 
$L_{\rm env} \propto M_{\rm env}^{3}$, where the envelop density 
distribution is taken as $\rho \propto r^{-2}$ and a solid body 
rotation is assumed to persist in the entire envelope. 
 Therefore, I do not expect HB stars, after losing $\sim 1/3$ of their 
 envelope on the RGB, to rotate at such high velocities. 
 Indeed, in order to account for the fastly rotating HB stars
Peterson Tarbell \& Carney (1983) already mentioned the 
possibility that planets can spin-up RGB stars.
 They cite the amount of the required angular momentum to be
about equal to the orbital angular momentum of Jupiter, which is $\sim 100$
times larger than that of the sun today. 
 The angular momentum problem on the HB is similar to that on the AGB, 
though it is less severe (Soker 1997; appendix below). 
\newline
{\bf (4)} In a recent paper Sosin {\it et al.} (1997) show that 
in the globular cluster NGC 2808 there are three subgroups in the blue HB.  
 Similar subgroups were found in the globular cluster M13 
(Ferraro {\it et al.} 1997). 
 Based on the stellar evolutionary simulations of Dorman {\it et al.} (1993)
Sosin {\it et al.} (1997) claim that the subgroups on the blue HB 
correspond stars having envelope masses of $M_{\rm {env}} \lae 0.01 M_\odot$, 
$0.02 \lae M_{\rm {env}} \lae 0.055 M_\odot$, 
and $0.065 \lae M_{\rm {env}} \lae 0.13 M_\odot$ from blue to red.
 The red HB stars are concentrated in the range  
$0.16 \lae M_{\rm {env}} \lae 0.22 M_\odot$.  
 They took the core mass to be $M_{\rm core} = 0.4847 M_\odot$. 
 The number of stars in each group, from red to blue, are 
$\sim 350$, 275, 70 and 60. 
What is the cause for the three subgroups in the blue HB found in several
globular clusters?
 
 I would like to suggest that interaction with planets on the RGB
can account for all these properties. 
 First, there are stars that will not interact on the RGB with any 
gas giant planet. 
 These stars will lose little mass, and they will form the red HB. 
 Terrestrial planets can influence the mass loss, but not by much, 
by increasing somewhat the angular momentum of the star. 
 The blue HB stars, I suggest, result from RGB stars that interact
with gas giant planets. 
 A planet entering the envelope of a RGB star releases both energy and
angular momentum, both of which are expected to increase the mass loss 
rate. 
Hence, the star reaches the HB with less massive envelope. 
  There are three evolutionary roots for star-planet systems 
(Livio \& Soker 1984): ($i$) evaporation of the planet in the envelope;
($ii$) collision of the planet with the core (i.e., the planets overflow its Roche 
lobe when at $\sim 1 R_\odot$ from the core); and ($iii$) expelling the
envelope while the planet survives the common envelope evolution. 
 These three roots may explain the three subgroups found by 
Sosin {\it et al.} (1997) in the blue HB of the globular cluster 2808
(Soker, in preparation). 
 Preliminary results suggest that the bluest subgroup, of very little 
mass, result from surviving planets or brown dwarfs of mass
$M_p \gae 10 M_J = 0.01 M_\odot$, where $M_J$ is Jupiter's mass. 
 The intermediate subgroup on the blue HB result from planets
that collide with the core. 
 They expel much of the envelope, but not all of it. 
 Their evaporation can adds some mass to the envelope, influencing the 
abundance.
 Some envelope mass must remains to ensure that friction and tidal forces
will force the planet to spiral-in and collide with the core. 
 Hence, this group of stars retain $M_{\rm env} \gae 0.02 M_\odot$. 
 This group is formed from planets of mass $3 M_J \lae M_p \lae 10 M_J$. 
 The third, and largest of the three blue HB subgroups, are stars in which 
planets were evaporated in their envelope. 
 These occur for lower mass planets $M_p \lae 3 M_J$, that (up to
a factor of a few) have escape velocity lower than the sound speed 
in the envelope at the location of evaporation. 
 They are of low mass, and evaporate at large radii (larger than 
the Roche lobe overflow separation of $\sim 1 R_\odot$), and hence release
much less gravitational energy than massive planets. 
 I find (work in preparation) that the gravitational energy released
goes as $\sim M_p^2$. 
 This relatively strong dependence on the planet's mass may explain the 
large separation between the intermediate and red subgroups of the blue HB. 

 Stellar binary mergers were suggested to account for the blue HB,
but Rood (1997) criticize this idea. 
 Rood's three comments against the {\it stellar}-binary scenario
do not hold against the star-planet scenario. 
(1) Planets do not change the general nature of the star beside the mass
loss, contrary to stellar companions which collide with the star. 
(2) We do not expect variation with location in the cluster, 
unlike in scenarios with binary collisions.  
(3) We do not expect the star-planet interaction to depend much on the
density of the cluster (beside influencing planetary system formation 
efficiency; see below), unlike for stellar collisions. 

 How does the planetary system scenario accounts for the different HB 
morphologies of different globular clusters?
 The different morphologies result both from the efficiency of planetary 
system formation and their properties, and from the evolution of 
stars on the RGB. 
 These factors depends on several other parameters:
\newline
{\bf (1) Metalicity:} (a) The metalicity influences the efficiency of 
planets formation. 
There is no god theory to predict the efficiency, but low
metalicity results in lower efficiency. 
 On the other hand, in globular cluster the HB stars result from 
main sequence stars fainter than the sun. 
 Fainter central stars evaporate less the pre-planetary disk, and
hence may allow Jovian planets to form more easily and closer to the star. 
(b) Metalicity determines the maximum radius which stars attain on the RGB,
being larger for metal rich stars. 
 Larger radii increase the chance of interaction with planets. 
\newline
{\bf (2) Global cluster properties:} The global properties of the cluster
(e.g., shape; density of stars; initial mass function) may determine 
the efficiency of planet formation. 
The globular clusters M13 and M3, for example, have many similar 
properties, but  M13 is more elliptical than M3. 
 M3 has no blue HB, while M13 has an extended blue HB. 
 What is interesting to the star-planet interaction scenario is that
there are more blue straggler in M3 than in M13 (Ferrao {\it et al.} 1997).
 This suggests that there are fewer stellar binary systems in M13. 
I would expect that if less stellar binary companions are formed,
then more planetary systems will form.  
 This might explain the anti-correlation of population on the
blue HB and the number of blue straggler stars observed in these
two globular clusters. 
\newline
{\bf (3) Age:} The age determines the initial mass (main sequence mass)
of the stars. 
 This influences both the envelope mass on the RGB, and 
the maximum radius on the RGB, being larger for less massive stars. 
 As mentioned above, the main sequence mass may determine the efficiency of 
planets formation as well. 

\clearpage 

\noindent {\bf{Appendix: The Failure of Single Star Models}}
\bigskip
   
 In a previous paper (Soker 1996b) I criticized the model of Asida and 
Tuchman (1995) on the ground that their model requires a binary companion 
to spin up the AGB envelope. 
 I showed that their extended envelope scenario for axisymmetrical 
mass loss on the AGB requires angular momentum on the AGB which is 
$\gae 2$ orders of magnitude from what a single star can supply. 
 The only possible source is a stellar binary companion of mass
$\gae 0.1 M_\odot$ at an orbital separation of several AU.
Even with the binary companion included, there are several other problems
in Asida \& Tuchman's model. 
 Effects due to extended envelopes on the AGB, though, deserve further study,
e.g., Harpaz {\it et al.} (1997) used it to propose periodic mass loss rate 
in a binary model. 

 Below I criticize two other models, mainly on the ground that they 
{\it must} incorporate a binary companion, stellar or substellar.
 The first is a scenario which assumes that the mass loss on the AGB
results from a radiation pressure on dust (Dorfi \& H\"ofner 1996). 
 The axisymmetrical mass loss result from imposing envelope
rotation. 
 Although it is quite plausible that rotation together with radiation 
pressure on dust form axisymmetrical mass loss, I claim that the 
model {\it must} incorporate a binary companion to spin-up the envelope.  
 In order for their proposed scenario to work, Dorfi \& H\"ofner (1996) 
require the angular velocity of their AGB star, of radius $R = 500 R_\odot$,
to be $\gae 10 \%$ of the Keplerian angular velocity. 
  We hardly find main sequence stars with such high rotational velocity; 
there is no way a single evolved AGB star can obtain this rotational 
velocity.
  Approximating the density profile on the AGB by $\rho \propto r^2$,
where $r$ is the radial distance from the center of the star, 
we find the envelope's moment of inertia to be 
$I_{\rm {env}} = (2/9) M_{\rm {env}} R^2$, where $M_{\rm {env}}$ is the
envelope's mass. 
  If we assume that the secondary orbital separation, before entering a   
common envelope phase and spinning up the envelope, was $\sim R$, 
we conclude that in order to spin-up the envelope as required by 
Dorfi \& H\"ofner the secondary mass should be 
$M_2 \gae 0.02 M_{\rm {env}}$, i.e., at least 10 times as massive as 
Jupiter for an envelope mass of $0.5 M_\odot$. 
 However, as Harpaz \& Soker (1994) show, the envelope's specific angular
momentum of an AGB star decreases with mass loss as 
$L_{\rm env}/M_{\rm env} \propto M_{\rm env}^{2}$. 
 Therefore, in order to supply the angular momentum for a longer time, 
the companion mass should be even higher than $\sim 0.01 M_\odot$. 
  Other effects that such a companion can cause (Soker 1997) should
then be considered as well. 

 The second model I find to have severe problems is that of Garcia-Segura 
{\it et al.} (1997), which is an extension of the model proposed 
by Chevalier \& Luo (1994; see also Chevalier 1995). 
 This model is based on the tension of the toroidal component of the 
magnetic field in the wind; the wind in the transition from the AGB to the
PN phase or the fast wind during the PN phase. 
 Close to the star the magnetic pressure and tension are negligible compared
with the ram pressure and thermal pressure of the wind. 
 As the wind hits the outer PN shell, which is the remnant of the slow wind,
it goes through a shock, slows down and the toroidal component 
of the magnetic field increases substantially. 
 This may result in the magnetic tension and pressure becoming the 
dominant forces.   
 In particular, the magnetic tension pulls toward the center and reduces 
the effective pressure in the equatorial plane. 
 According to this model (Chevalier \& Luo 1994; Garcia-Segura {\it et al.}
1997), then, the equatorial plane will be narrow, leading to an elliptical 
or bipolar PN. 
 I find four main problems in the model described above.  
\newline
{\bf (1) The Energy Source of the Wind:} 
 The efficiency of this model is determine by a parameter given by
(Chevalier \& Luo 1994; Garcia-Segura {\it et al.}
$\sigma=(B_s^2r_s^2/\dot M_w v_w)(v_{\rm {rot}}/v_w)^2$, 
where $B_s$ is magnetic field intensity on the stellar surface,
$r_s$ the stellar radius,  $\dot M_w$ the mass loss rate into the wind, 
$v_w$ the terminal wind velocity and $v_{\rm {rot}}$ the equatorial 
rotational velocity on the stellar surface.
 Using the expression for the magnetic energy luminosity  
 $\dot E_B=4 \pi r_s^2 v_w B_s^2 / 8 \pi$ and for the kinetic energy 
luminosity $\dot E_k = \dot M_w v_w^2/2$ we can express $\sigma$ as  
\begin{equation}
\sigma= {{\dot E_B}\over{\dot E_k}} 
\left({{v_{\rm {rot}}}\over{v_w}} \right)^2. 
\end{equation}
 For the model to be effective it is required that $\sigma \gae 10^{-4}$,
but a typical value of $\sigma \simeq 0.01$ is used by 
Garcia-Segura {\it et al.} (1997). 
 For the sun $\sigma \simeq 0.01$ and 
$(v_{\rm {rot}}/v_w)^2 \simeq 2 ,\times 10^{-5}$ (Chevalier \& Luo 1995). 
 However, in the sun it is magnetic activity which determines the mass loss
rate, as we see from the ratio $\dot E_B/\dot E_k \simeq 500$.  
 It is commonly assumed that radiation pressure derives the winds of
central stars of PNs, and that pulsation together with radiation pressure
derives the wind of AGB stars, and red giants in general. 
 Therefore, the sun is not a good example of this model for singly evolved 
stars. 
 Soker \& Harpaz (1992) argue that dynamo activity might produce strong 
enough magnetic fields in AGB stars only if a binary companion 
(stellar or substellar) spins-up the envelope.  
 Such effect will increase the ratio $(v_{\rm {rot}}/v_w)$ as well. 

\noindent {\bf (2) Angular Momentum:} 
 If magnetic energy does not derive the wind then 
$\dot E_B/\dot E_k \lae 1$, and the model of magnetic shaping requires 
$ v_{\rm {rot}}/v_w \gae 0.01$.  
 Such rotation velocity is {\it impossible} for singly evolved AGB or
post-AGB stars to attend (Harpaz \& Soker 1994). 
 Therefore, spinning-up by a binary companion {\it must} occur even if the
wind does not result from magnetic activity. 
 
 From problems (1) and (2) it is clear that in order for the magnetic shaping
to be of any significance, a {\it substantial} spinning-up by a binary
companion is required (see also Livio 1995). 
 But even if this condition is met, I find two other problems with this 
model concerning shaping on a large scale.
 The two problems result from MHD instabilities. 
 Other MHD instabilities might exist as well (Livio 1995). 

\noindent {\bf (3) The Shape of the Ejected Magnetic Field:} 
 The magnetic shaping model requires that the magnetic field lines
will circle the central star in the equatorial plane. 
 However, it is not clear that this will be the case. 
 Because of MHD instability the magnetic filed escape from the sun 
in non-axisymmatrical magnetic flux loops (e.g., Bieber \& Rust 1995; 
Caligari, Moreno-Insertis, \& Sch\"ussler 1995). 
 Many flux loops which escape the sun  do not circle the sun. 
 It is possible, though, that when the magnetic pressure is low then 
more flux loops will circle the star. 
But then, as I showed above, the angular momentum problem is severe. 

\noindent {\bf (4) Reconnection inside the Hot Bubble:}
 For magnetic field to be of any significance, it should be regenerated
by a stellar dynamo. 
 The idealized toroidal magnetic field that result from a dynamo has 
opposite directions in the two stellar hemispheres (e.g., Bieber \& Rust 1995).
 Therefore, as the magnetic pressure becomes dominate after the wind
slows down, I expect that reconnection of magnetic field lines close 
to the equatorial plane will occur. 
 This will decrease the magnetic pressure and tension near the equatorial 
plane, and as a consequence will reduce the efficiency of the model. 
 
 The last two problems relates only to the large scale shaping.
 Magnetic filed may still be strong but with a short coherence length. 
  I think that when substantial spinning occurs and if dynamo activity 
becomes efficient, magnetic fields may play a substantial role on small 
scale shaping, e.g., MHD instability modes on small scales.

\end{document}